# Patterns of mast fruiting - a stochastic approach

Ciprian Palaghianu, Marian Drăgoi

Abstract
Mast fruiting represents a synchronous population behaviour which can spread on large landscape areas. This reproductive pattern is generally perceived as a synchronous periodic production of large seed crops and has a significant practical importance to forest natural regeneration in order to synchronize cuttings. The mechanisms of masting are still argued and models of this phenomenon are uncommon, so a stochastic approach can cast significant light on some particular aspects. Trees manage to get synchronized and coordinate their reproductive routines. But is it possible that trees get synchronized by chance, absolutely random? Using a Monte Carlo simulation of seeding years and a theoretical masting pattern, a stochastic analysis is performed in order to assess the chance of random mast fruiting. Two populations of 100 trees, with different fruiting periodicity of 2-3 years and 4-6 years, were set and the fruition dynamic was simulated for 100 years. The results show that periodicity itself cannot induce by chance the masting effect, but periodicity mathematically influences the reproductive pattern.

Introduction
Mast fruiting or masting represents a reproductive pattern of some tree species which exhibit a rhythm of fruiting and set abundant crops only in certain years, with a more or less regular periodicity. This phenomenon is truly remarkable in forest ecosystems and its effect spreads and influences a great variety of other species – from insects and birds to rodents and ungulates (Vander Wall, 2001).
Of course, from the forester's point of view, the variability of trees' fruition represents a challenge to natural regeneration of forests, so masting has also a practical significance, considering the effort of synchronizing cuttings with the seeding years.
There are some key aspects regarding the masting years: the variability of crops and the periodicity, two elements that can be related to individual trees. However, the mast phenomenon is not an individual feature, but a group behaviour. We may assume that trees manage to synchronize and coordinate their reproductive actions. So the third key element in masting is synchronicity.
This phenomenon has been considered to be intensely influenced by the environmental conditions and resources availability (resource matching hypothesis), but there are studies that offer physiological and/or ecological explanation, presenting this reproductive strategy as an adaptation of tree species to seeds predators (predator satiation hypothesis) or as an evolved way to increase pollination success (pollination efficiency hypothesis).

The mechanisms of masting are arguable and certainly still unclear, but these hypotheses represent valuable starting points.
The resource matching hypothesis (Busgen & Munch 1929; Sork 1993; Kelly 1994) represents one of the oldest and most straightforward ways of explaining the variability of seed production. More recent studies conducted at global scale (Koenig & Knops, 2000, 2005) partially confirmed this hypothesis, underlining that environmental disturbances could trigger similar reproductive patterns. Koenig (2002) emphasizes the influence of the Moran effect on mast seeding, showing that environmental synchrony can induce similarity in tree populations' behaviour. However, most authors consider that there are supplementary factors that influence the masting process.
The pollination efficiency hypothesis represents a rather different approach in explaining the evolution of masting patterns (Janzen, 1967; Norton and Kelly 1988; Smith et al. 1990; Kelly et al. 2001). This assumption presumes that intermittent and coordinated peaks in fruiting might increase the pollination success.
Another consistent explanation is suggested by the predator satiation hypothesis (Janzen 1971; Silvertown, 1980; Augspurger, 1981) which assumes that discontinuous abundant seed production can lead to lower seed predation in masting years.
More recent theories propose that masting phenomenon is related to climate changes, ENSO cycles or solar activity (Koenig, 1999; Williamson & Ickes, 2002; Piovesan & Adams, 2005; Pearse et al., 2014), opening new perspectives in clarifying the masting mechanisms.

There are still ample debates regarding the evolution of this reproductive pattern, but the main hypotheses are focusing on the variability of crops and its periodicity, leaving the synchronicity aspect behind (Kelly & Sork, 2002). But there are consistent differences between the periodicity and synchrony of abundant fruiting. Early studies in phenology, flowering and reproductive synchrony (Janzen, 1967, 1978; Waller, 1979, Silvertown, 1980; Augspurger, 1981, 1983) did not elucidate the drivers of synchrony.

Periodicity is related to certain cycles, but masting involves synchronized cycles. Why do different trees flower and fruit at the same time? One important aspect that must be clarified is not why trees have variable crops as individuals, but how they get synchronized on large areas and manage to coordinate their reproductive behaviour. What is causing the synchrony?

One coherent and plausible explanation leads to the Moran effect (Ranta et al., 1997; Koenig, 2002), the idea that environment can induce synchronous fluctuation in population size. However, this effect is based on resources similarity and it can be very effective regarding population size, but it cannot be very sensitive to population behaviour.

One thought-provoking approach could use as starting point the possibility that trees get synchronized by chance, absolutely at random. Obviously, it does not seem a reasonable premise, but an arguable problem requires sometimes a questionable approach. However, given a specific periodicity for a tree population with individuals in random states of fruition, what kind of mast patterns would generate a simulation of fruition dynamic?

The mechanisms of masting are certainly still unclear and models of this phenomenon are not common, so a stochastic approach can cast significant light on some certain aspects.

Material and methods

Essentially the idea of the paper follows the reproductive dynamic of two populations that are unsynchronised at individual level regarding their fruition. The uncertainty and randomness is partially introduced in the model of achieving abundant fruition and the goal is to assess if pure chance can lead a population from unsynchronised state to a synchronised one.

The two hypothetical populations have different periodicity of fruition in order to test the potential influence of periodicity on masting. This is the element we wanted to decipher in this paper, so we suggest an alternative stochastic analysis for assessing fruition patterns, based on random individual data.

The fruition is represented as a stochastic process, using randomly generated hypothetical tree populations, with the same periodicity and different states of fruition.

The mast years were regarded as simulated events for each individual tree, and the fruition dynamic was analysed for both populations. Using Monte Carlo simulations of seeding years we set this scenario for two populations of 100 trees, with different fruiting periodicity of 2-3 years and 4-6 years. Then we simulate the fruition dynamic for 100 years.

The starting point was represented by the generation of a randomly initial state for all trees of the two populations. The initial state was created using the true random number service (via atmospheric noise) offered by the website www.random.org. There were generated 20 sets of 100 individual data for each population in order to achieve a 95% confidence envelope. The individual value for each tree that marked the initial state was an integer number which renders the number of years past from the last abundant fruition (mast year). It was also allowed the random generation of an individual 0/zero state that marks a tree which is experiencing an abundant fruition year.

Afterward all data were included in a probabilistic simulation of fruition for a period of 100 years. The model used to achieve the next masting year was based on a probabilistic function created in Visual Basic for Applications and the data were processed in Microsoft Excel.

The function increments each individual value, unless a masting event is occurring, in which case the value resets to zero. The simulated masting event occurs with a variable probability, depending on the years past from the previous masting and the population fruiting periodicity.

There were used two different probabilistic fruition patterns for the two populations:
- for the population with a 2-3 years periodicity, the next masting year will occur with a 15% probability in 1st year, 65% in 2nd year, 95% in 3rd year and with 99% after that.
- for the population with a 4-6 years periodicity the next masting year will occur with a 1% probability in 1st year, 5% in 2nd year, 15% in 3rd year, 60% in 4th year, 80% in 5th year, 95% in 6th year and with 99% after that.

In this way the general population fruiting periodicity is preserved for each individual tree, but the stochastic model is flexible enough to allow individuals to randomly adjust their periodicity within some certain limits.

These aleatory adjustments can lead to changes in the population's degree of fruiting synchronicity.

The recurrent process of fruiting simulation was applied for a period of 100 years for all the trees within all the 20 sets of each population, resulting in 400,000 simulated values.

The data from the 20 simulated sets were used to compute the average number of masting tree per years and the 95% confidence envelope of this multiple event, for each of the two populations.

For model manageability there were made several simplifying assumptions:
- homogenous environmental conditions (including the lightning conditions)
- all trees are mature and have the capability to fruit
- for the analysed time period the mortality did not affect the tree populations
- age of trees neither affect the periodicity, nor the synchronization of fruition
- only the mast years are considered events that were used for assessing fruition synchrony (the other years are not events regardless of their fruition intensity).

Results and Discussion

For all the 20 simulated sets of data of each population and for all the 100 years it was computed the percent of the trees with a masting state. Subsequently, for every year there were calculated the basic statistic indicators, such as minimum, maximum, average and coefficient of variation.

Descriptive statistics allowed us to compute the indicators that described the characteristics of the two populations regarding the reproductive pattern. The average percentage of masting trees in the populations with a 2-3 years periodicity (31.66%) is much higher than the same indicator for the population with a 4-6 years periodicity (19.62%),

There are significant differences between the means of the two populations with different fruiting periodicity, as shown in Table 1.

It is obvious that periodicity influences the way that trees from a population synchronize their fruition. Higher periodicities lead to smaller values of the percentage of masting trees and to higher coefficients of variation.

Basically, from the mathematically point of view, this fact means that species with a higher periodicity of fruition are forced to intensify their efforts and spend more energy in order to achieve the status of masting. It's easier, statistically speaking, for a population with a lower periodicity to synchronize the abundant seed crops of its individuals.

Due to these differences regarding the percentage of masting trees, a mast year should be considered in a different manner for each species. It should not be regarded as linked to a static percentage, but as a function of species' periodicity. Thus, the use of different scales should be more appropriate and equitable.

Using the annual averages, the results of the simulations were plotted in individual charts and the absolute minimum and maximum values from the 20 simulated sets were used to generate the 95% confidence envelope of this multiple event, for each of the two populations.

The two charts (Figure 1 and 2) present the average number of masting tree per year and the confidence envelope for the two considered populations with a 2-3 years periodicity, respectively a 4-6 years periodicity, during a period of 100 years.

The graphic representations emphasize the differences between the two populations. Each population has distinct reactions and the small adjustments of individual fruiting periodicity that lead to group synchronization are becoming visible mostly in the first 20 to 25 years from the initial moment of fruiting simulation. This first period shows a better synchronization of the reproductive process, and the charts show higher peaks, but then uniformity becomes more evident.

This particular behaviour can be explained by the influence of the initial state on synchronization. The random initial state might influence the results and the impact could be observed in both charts for the initial decades.

For example, in a previous test, the initial state did not contain any tree in masting phase and as a result the following years were characterized by very high degrees of masting trees. In this situation, the average degree of masting for the first year of the simulation was nearly 20% higher in the case of populations with a 4-6 years periodicity and nearly 40% higher in the case of populations with a 2-3 years periodicity. Actually, any delay of the fruition process simulated by a year with no trees in masting state, increases the chance of masting in the next year.

Regarding the starting point of the simulation, it is possible that a truly random initial state not to be the best suited option. This choice can be reconsidered and different types of initial distributions could be tested and used. Furthermore, the initial state could be calibrated using real data to assess the percentage of masting trees in a population with a similar fruiting periodicity.

Moving on to another issue, studying the fruiting charts of individual populations, the influence of periodicity gets more evident. A better synchronization of fruition is observed (see Figure 3), compared with the chart that plots the average degree of masting trees. In some way it is normal because using an average built on 20 simulated sets, the synchrony trend gets obscured by the smoothing effect of the mean.

The amplitude between mast years and regular year is better marked and the masting peaks generally surpass the mean by more than one standard deviation. However, the masting percentage is still below 50%. There are recorded exceptional years but not enough to be generally considered masting years.

Tab. 1: The t test for the two populations

|  | 2-3 yrs. periodicity | 4-6 yrs. periodicity |
|---|---|---|
| Mean | 31.66 | 19.62 |
| Variance | 2.68 | 5.12 |
| Observations | 100 | 100 |
| df | 180 |  |
| t Stat | 43.12 |  |
| P(T<=t) two-tail | 8.139E-97 |  |
| t Critical two-tail | 1.97 |  |

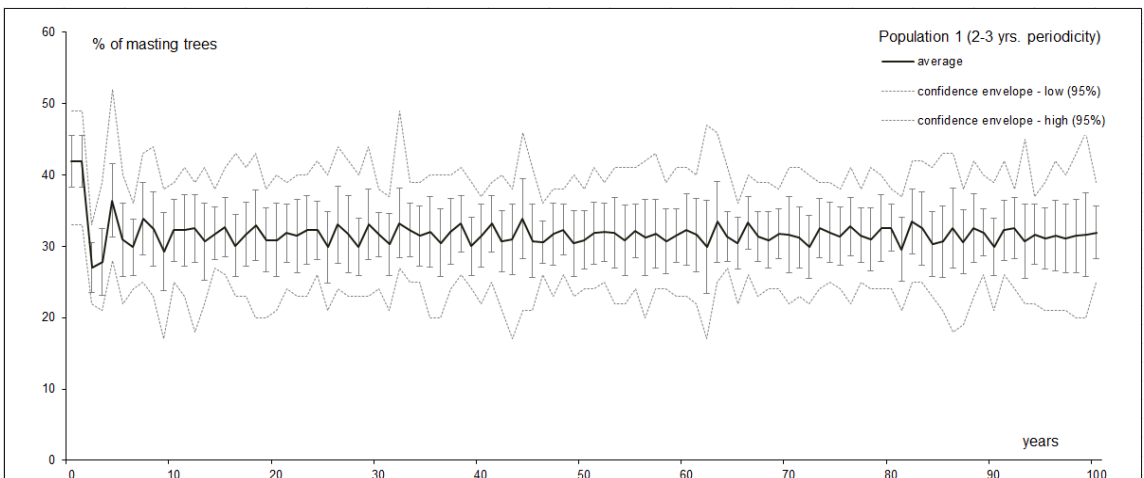

Fig. 1: The mast fruiting dynamic (2-3 yrs.per.)

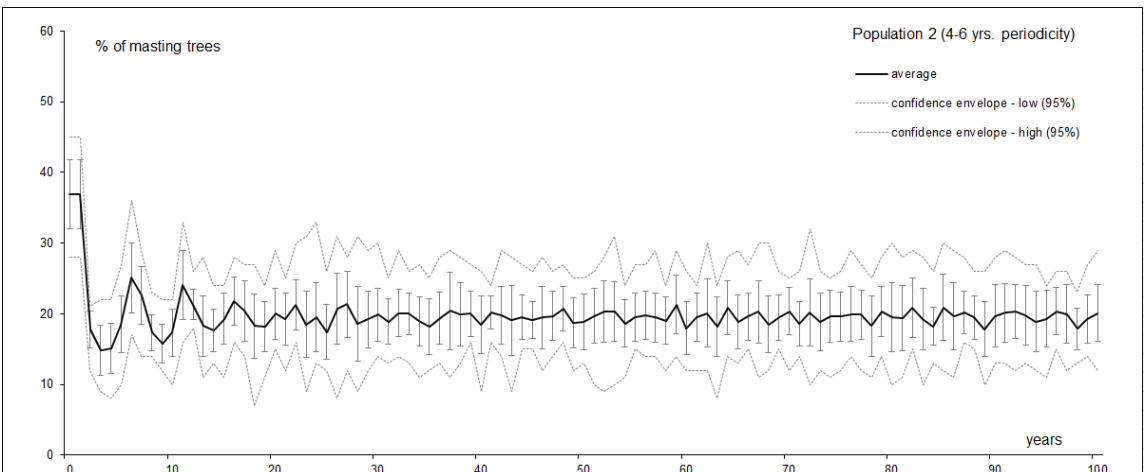

Fig. 2: The mast fruiting dynamic (4-6 yrs.per.)

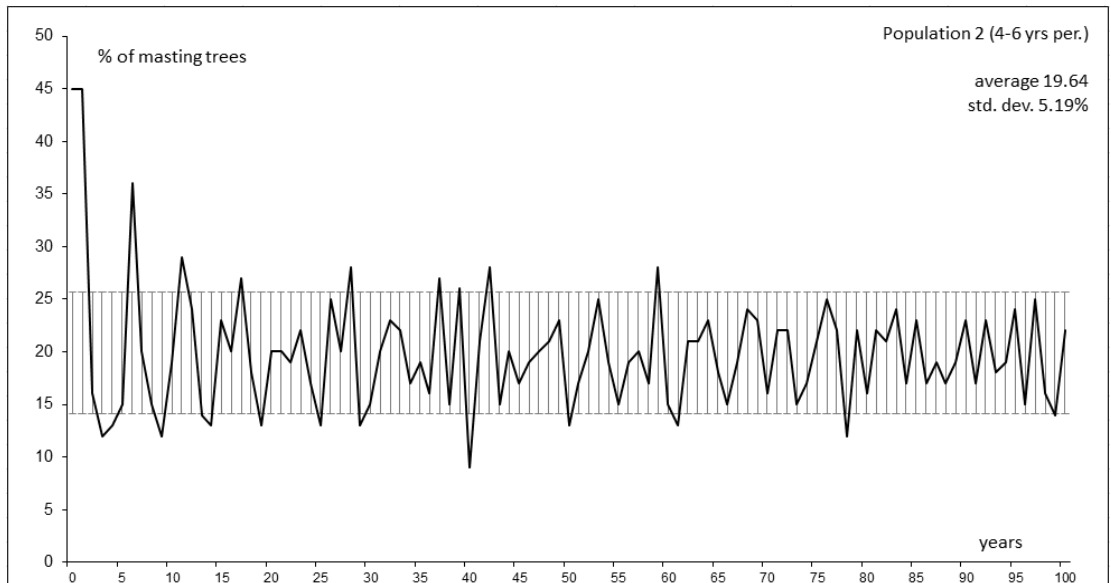

Fig. 3: Individual simulation of a population

## Conclusion

The patterns of mast fruiting continue to puzzle researchers and various methods of investigation are now used. Recent studies successfully linked masting with phenology or dendrochronology, some trendy approaches in the context of climate changes (Drobyshev et al, 2010). It is clear that it is necessary to focus more on historical records of fruition and good results were obtained for several species. Latest studies successfully reconstructed fruiting series for beech that extend to more than two centuries (Drobyshev et al, 2014). Such results bring fresh data not only related to periodicity but could allow synchronization tests of fruiting on larger areas.

There are a few issues regarding the mast fruiting study that must be improved. One particular aspect that must be solved in the near future is finding a better way of evaluating fruiting synchronicity. Another aspect that needs to be clarified is represented by the reconsideration of masting years from the statistical point of view. A reasonable way is considering a masting event only if the percentage of abundant fruiting trees exceeds, for example, one or two standard deviations of this parameter when taking into account a standard period of time.

Regarding the results of the paper, we are aware that the model we presented is only a draft version, but at this point we have already tested some reproductive patterns, made thousands of iterations and we think the final model could prove feasible and flexible. It is clear that periodicity itself cannot induce by chance the masting, but evidently, periodicity mathematically can influence synchronicity. It is harder for species with high values of fruiting periodicity to synchronize the abundant seed crops of its individuals.

There must be other factors that induce synchronous fluctuation in fruition, and we believe they could be integrated as probabilities in the final model. Maybe we should take into account even some various cycles in nature which can influence the fruiting periodicity and finally causing synchrony.

These results confirm the necessity of focusing more intensively on deciphering the intrinsic mechanisms of masting.

Author's contact
Eng. Ciprian Palaghianu, Ph.D.,
Eng. Marian Drăgoi, Ph.D.,
Department of Silviculture and Environmental Protection, Forestry Faculty,
Stefan cel Mare University of Suceava
Universitatii Street, 13, Suceava, Romania
Tel: (+40) 0745 614 487
Email:    cpalaghianu@usv.ro,  dragoi@usv.ro